\documentclass{article}

\usepackage{arxiv}

\usepackage[utf8]{inputenc} % allow utf-8 input
\usepackage[T1]{fontenc}    % use 8-bit T1 fonts
\usepackage{hyperref}       % hyperlinks
\usepackage{url}            % simple URL typesetting
\usepackage{booktabs}       % professional-quality tables
\usepackage{amsfonts}       % blackboard math symbols
\usepackage{nicefrac}       % compact symbols for 1/2, etc.
\usepackage{microtype}      % microtypography
\usepackage{lipsum}		% Can be removed after putting your text content
\usepackage{graphicx}
\usepackage{natbib}
\usepackage{doi}
\usepackage{siunitx}
\title{Stress Detection from Multimodal Wearable Sensor Data}

\author{ \href{https://orcid.org/0009-0009-5585-2183}{\includegraphics[scale=0.06]{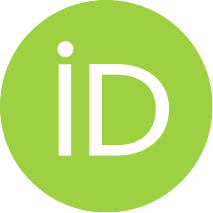}\hspace{1mm}Paul Schreiber}
\\
Department of Data Engineering\\
Helmut-Schmidt University\\
Hamburg, Germany \\
\texttt{schreibp@hsu-hh.de} \\
\And
\href{https://orcid.org/0009-0006-3617-0134}{\includegraphics[scale=0.06]{orcid.pdf}\hspace{1mm}Beyza Cinar} \\
Department of Data Engineering\\
Helmut-Schmidt University\\
Hamburg, Germany \\
\texttt{cinarb@hsu-hh.de} \\
\And
\href{https://orcid.org/0009-0006-7138-7902}
{\includegraphics[scale=0.06]{orcid.pdf}\hspace{1mm}Lennart Mackert} \\
Department of Data Engineering\\
Helmut-Schmidt University\\
Hamburg, Germany \\
\texttt{lennart.mackert@hsu-hh.de} \\
\And
\href{https://orcid.org/0000-0003-3458-4748}{\includegraphics[scale=0.06]{orcid.pdf}\hspace{1mm}Maria Maleshkova} \\
Department of Data Engineering\\
Helmut-Schmidt University\\
Hamburg, Germany \\
\texttt{maleshkm@hsu-hh.de} \\
}

\renewcommand{\shorttitle}{\textit{arXiv} Template}

\hypersetup{
pdftitle={A template for the arxiv style},
pdfsubject={q-bio.NC, q-bio.QM},
pdfauthor={David S.~Hippocampus, Elias D.~Striatum},
pdfkeywords={First keyword, Second keyword, More},
}

\begin{document}
\maketitle

\begin{abstract}
Human-Computer Interaction (HCI) is a multi-modal, interdisciplinary field focused on designing, studying, and improving the interactions between people and computer systems. This involves the design of systems that can recognize, interpret, and respond to human emotions or stress. Developing systems to monitor and react to stressful events can help prevent severe health implications caused by long-term stress exposure. Currently, the publicly available datasets and standardized protocols for data collection in this domain are limited. Therefore, we introduce a multi-modal dataset intended for wearable affective computing research, specifically the development of automated stress recognition systems. 
We systematically review the publicly available datasets recorded in controlled laboratory settings. Based on a proposed framework for the standardization of stress experiments and data collection, we collect physiological and motion signals from wearable devices (e.g., electrodermal activity, photoplethysmography, three-axis accelerometer). During the experimental protocol, we differentiate between the following four affective/activity states: neutral, physical, cognitive stress, and socio-evaluative stress. These different phases are meticulously labeled, allowing for detailed analysis and reconstruction of each experiment. Meta-data such as body positions, locations, and rest phases are included as further annotations. In addition, we collect psychological self-assessments after each stressor to evaluate subjects' affective states. The contributions of this paper are twofold: 1) a novel multi-modal, publicly available dataset for automated stress recognition, and 2) a benchmark for stress detection with 89\% in a binary classification (baseline vs. stress) and 82\% in a multi-class classification (baseline vs. stress vs. physical exercise).
\end{abstract}

% keywords can be removed

\section{Introduction}
\label{sec:introduction}
Stress is linked to adverse effects on mental health, and may worsen symptoms of those suffering from migraines and stress disorders \citep{hovsepian2015cstress}. Especially the long-term exposure to stress may be detrimental to health and even cause physiological issues such as heart disease \citep{miller2006turning}. Commonly employed methods to assess stress include psychological self-reports (e.g., Perceived Stress Scale \citep{cohen1994perceived}. These types of measures, however, are rather impractical in everyday life, subjective, and not continuous, limiting the monitoring of stress exposure. To address these issues, researchers have employed non-invasive wearable devices for continuous monitoring of physiological signals, which can be used to develop automated stress recognition systems \citep{plarre2011continuous, hovsepian2015cstress, gjoreski2016continuous}. Accordingly, stress detection has been studied in different settings, including data from laboratory studies (e.g., \citep{benchekroun2023cross, schmidt2018introducing, gjoreski2016continuous, birjandtalab2016non, mishra2020continuous, plarre2011continuous, mozos2017stress, parent2020pass} and field studies (e.g., \citep{mishra2020continuous, hosseini2022multimodal, schneegass2013data, taylor2013warwick, hovsepian2015cstress, mundnich2020tiles, yau2022tiles} to study subjects' exposure to different stressors. These studies achieve promising results in predicting or classifying the exposure to stress stimuli, and the resulting datasets are indispensable to the development of automated stress recognition systems. However, the domain needs broader validation of these results and testing whether the systems employed generalize well to other samples.

The lack of validated and generalizable stress management systems can be attributed to number of reasons. First of all, publicly available datasets in the domain of stress detection are rare, as pointed out by \cite{benchekroun2023cross}.
Secondly, data collection and experimental setups in stress research and affective computing still lack standardized procedures that facilitate structural unity of datasets \citep{schreiber2024trrraced}. However, a standardized protocol to infer human emotions \citep{miranda2017meaningful}, and other affective states such as stress, would vastly benefit the development of reproducible and generalizable recognition systems. Furthermore, the datasets used to develop automated stress recognition systems often differ in sensor technologies, stress elicitation methods, and vital signs recorded, restricting the comparability of results (compare \citep{mishra2020evaluating}). 
Lastly, and related to the previous issues, is the lack of proper annotation of the datasets. Datasets for stress recognition are often complex and characterized by their temporal, continuous, and multi-modal structure. Adequate labeling and annotation of such datasets is indispensable, specifically to facilitate the validation of research results by other researchers and to integrate data with other datasets.

\sloppy{
This paper introduces a novel multi-modal dataset, thus aiding the development of more robust stress recognition systems. VitaStress constitutes a novel dataset in the domain of affective computing, particularly automated stress recognition, including different types of stressors and a neutral baseline. 
During controlled laboratory experiments, we follow the TRRRACED-framework (\textbf{T}owards \textbf{R}eproducible, \textbf{R}eplicable and \textbf{R}e-suable \textbf{A}ffective \textbf{C}omputing \textbf{E}xperiments and \textbf{D}ata), proposed by \cite{schreiber2024trrraced}. The framework aims to introduce a standardized protocol for stress experiments, including annotations and metadata to avoid the aforementioned issues. By applying this framework, the present research paper aims to promote standard protocols that facilitate the structural unity across datasets, increasing comparability among datasets and study results. Furthermore, the dataset contains processed vital and kinematic data such as heart rate, intervals between two consecutive heartbeats (RR-interval), temperature, respiration, skin conductance, as well as raw data of photoplethysmography (PPG), skin conductance, and a three-axis accelerometer. In addition to a neutral baseline recording, three types of moderate stress were evoked in subjects: physical, cognitive, and socio-evaluative stress. We also provide a comprehensive overview and systematic analysis of publicly available stress datasets. The contributions of this paper are threefold:  1) a systematic review of publicly available stress recognition datasets, and 2) a novel multi-modal, publicly available dataset for automated stress recognition and 3) validation of a proposed framework for standardization. 
We expect the dataset and provided summary statistics to bear the potential for more comprehensive analysis and better stress recognition results in future research.}

This paper is structured as follows: Section \ref{sec:related_work} discusses commonly used emotion and stress momdels, as well as the available datasets in automated stress recognition. Section \ref{sec:data_and_methods} presents the experimental set-up and data collection. Section \ref{sec:vitastress} presents the VitaStress dataset. Section \ref{sec:results} shows how the dataset was evaluated for two different classification tasks. Next, section \ref{sec:discussion} discusses the relevance of the results of this research paper. Finally, section \ref{sec:conclusion} concludes this research.

\section{Related Work}
\label{sec:related_work}

\subsection{Emotion and Stress Models}
\label{subsec:emotion_and_stress_models}
Two types of emotional models are frequently employed to operationalize stress in the Affective Computing and Stress recognition literature: categorical and dimensional models. Categorical models represent emotions and affect in discrete categories. Regarding categorical models, different theories distinguish between \textit{basic} emotions. For example, \cite{ekman1978facial} distinguish between \textit{joy, sadness, anger, fear, disgust, and surprise}. Accordingly, it is argued that these emotions are discrete and are represented by different physiological responses. Plutchik developed another classification \citep{plutchik1980emotion}, which introduced the 'wheel of emotions', comprising eight basic emotions: \textit{grief, amazement, terror, admiration, ecstasy, vigilance, rage, and loathing}. These emotions can be expressed at different intensity levels and blended. At the fundamental level, categorical models are confined to the predefined emotion categories and do not give insight into the intensity of the emotion felt. The choice of the basic emotions is somewhat arbitrary but may not align across cultures or individuals \citep{schmidt2019wearable}.

\begin{figure}
\centering
\includegraphics[width=0.6\columnwidth]{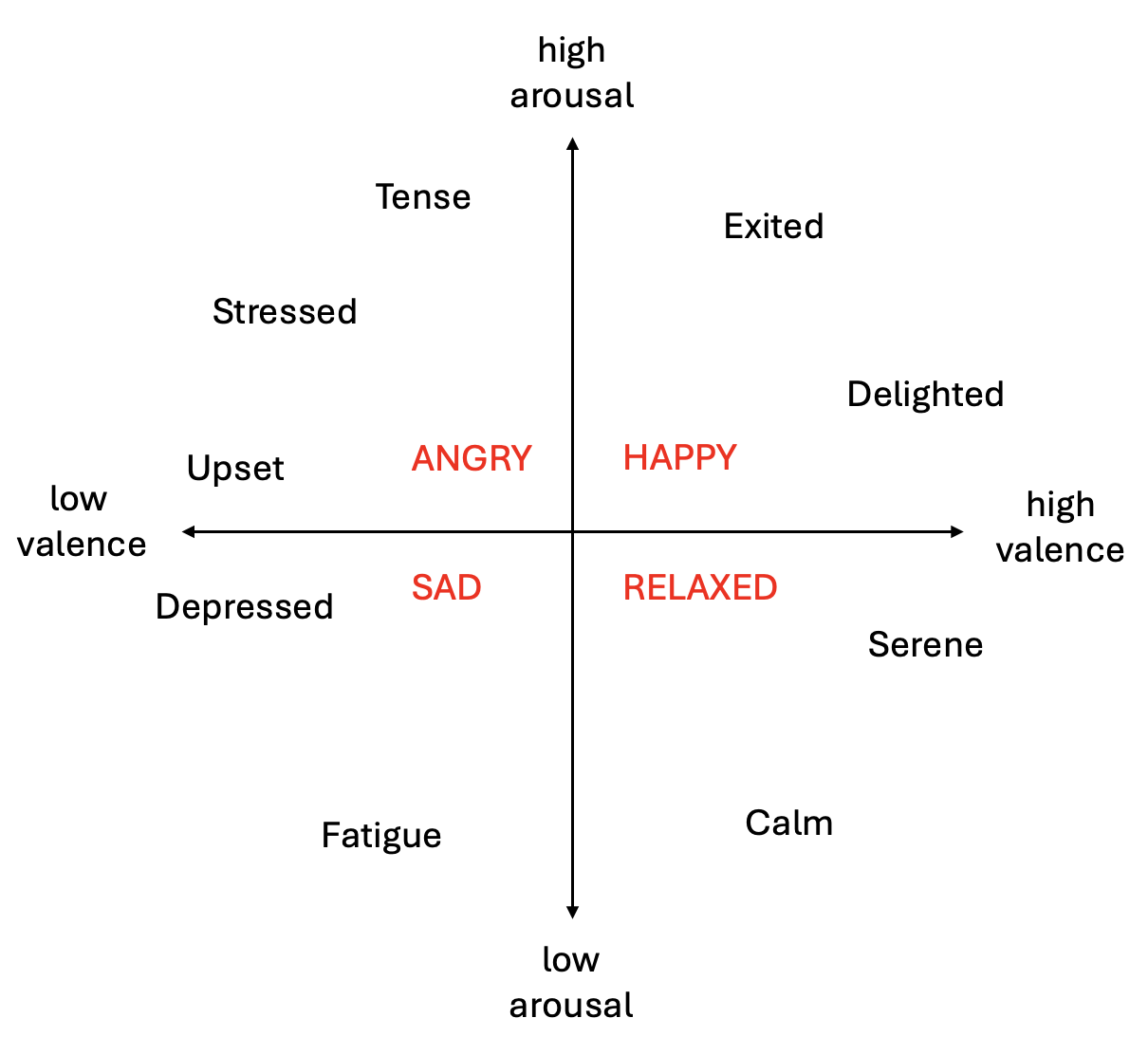}
\caption{Adaption of the circumplex model from \cite{schmidt2019wearable}} \label{circumplex}
\end{figure}

For dimensional models, affective states are represented as discrete points mapped into a multidimensional space. The most notable dimensional model that has been often employed in the affective computing domain (e.g., \citep{sharma2019dataset, markova2019clas}) is the circumplex model by \citep{russell1980circumplex}, also known as the valence and arousal model. The circumplex model represents affective states along two dimensions, namely \textit{valence} and \textit{arousal}. On one hand, the valence axis indicates the perception of how positive or negative the current affective state is. 
On the other hand, the arousal axis shows the activation level, whether one feels energized or exhausted. The four quadrants of the circumplex model allow for broadly categorizing emotions and affective states into \textit{sad, relaxed, angry, happy}, which can be further classified depending on the level of intensity of each axis. As shown in Figure \ref{circumplex}, the circumplex model allows for modeling different intensities of low arousal/low valence (LALV), low arousal, high valence (LAHV), high arousal/low valence (HALV), and high arousal/high valence \citep{schmidt2019wearable}. For example, stress is in the high arousal/low valence area. The circumplex model can be extended, usually with an axis for \textit{dominance} to determine the perceived control over the evaluated situation.

\subsubsection{Publicly Available Datasets}
\label{subsec:publicly_available_datasets}

Automated stress detection has been addressed by many studies, either in controlled laboratory settings (e.g. \cite{benchekroun2023cross, schmidt2018introducing, gjoreski2016continuous, birjandtalab2016non, mishra2020continuous, plarre2011continuous, mozos2017stress, parent2020pass}), or in real-life scenarios (e.g. \cite{hosseini2022multimodal, schneegass2013data, taylor2013warwick, hovsepian2015cstress, mundnich2020tiles, yau2022tiles}). Studies in this domain usually focus on collecting datasets in either of the aforementioned settings, to use this data in downstream tasks such as stress detection, formulated as a supervised or unsupervised learning task. Data collection plays a pivotal role in such studies due to the limited number of public datasets. Table \ref{tab:my-table} provides an overview of the public datasets in the domain of stress detection, presented to the best of our knowledge. This overview includes datasets recorded in laboratory settings or real-life scenarios, yet the present research paper mainly focuses on data collected in controlled laboratory settings.

\begin{table}
\centering
\resizebox{\columnwidth}{!}{%
\begin{tabular}{llll}
Authors & Year  & Dataset  & Context   \\ \cline{1-4} 
\cite{schmidt2018introducing} & 2018 & WESAD & Laboratory \\
\cite{birjandtalab2016non} & 2016 & Non-EEG & Laboratory \\
\cite{iqbal2022stress} & 2022 & Stress-Predict  & Laboratory \\
\cite{parent2020pass}& 2020 & PASS & Laboratory \\
\cite{markova2019clas}& 2019 & CLAS & Laboratory \\
\cite{hongn2025wearable} & 2025 & Wearable Dataset & Laboratory \\
\cite{koldijk2014swell} & 2014 & SWELL-KW & Knowledge Work\\
\cite{schneegass2013data} & 2013 & Driver Workload & Driving \\
\cite{taylor2013warwick} & 2013 & Driving Monitoring Dataset (DMD) & Driving \\
\cite{mundnich2020tiles} & 2020 & TILES-18 & Hospital Work \\
\cite{yau2022tiles} & 2022 & TILES-19 & Hospital Work \\
\cite{hosseini2022multimodal} & 2022 & Physiological Stress Signals & Hospital Work \\
\end{tabular}%
}
\caption{Overview of publicly available datasets in the context of stress recognition}
\label{tab:my-table}
\end{table}

Lab-based stress studies focus on controlled interventions to induce (different kinds of) stress in subjects. Usually, sequences of several stressors are employed to elicit physical, cognitive, or socio-evaluative stress (e.g., \citep{mishra2020continuous, gjoreski2016continuous}. \cite{parent2020pass}) present a stressor which is a combination of physical and cognitive tasks at the same time. Compared to field studies, controlled interventions allow for detailed annotations and labeling. 

Various methods have been developed to elicit the intended type of stress in subjects. Physical stress has been induced either by immersing the hand of a subject in ice-cold water \citep{plarre2011continuous, mishra2020continuous, mishra2020evaluating} or during physical exercise \citep{ birjandtalab2016non, hongn2025wearable}. Cognitive stress is usually induced by employing a mental arithmetic task, as seen in the Trier Social Stress Test (TSST) \citep{kirschbaum1993trier}, the Stroop Color and Word Test \citep{benchekroun2023cross}, or even video games \cite{parent2020pass}. Socio-evaluative stress is elicited through a semi-public speaking task \citep{kirschbaum1993trier}, during which subjects present a task before a jury. The public speaking task is conducted under the pretense of evaluation by an expert and is also part of the TSST. Considering the many studies and applications in automated stress detection systems, it becomes apparent that standardized protocols such as the TSST are reliable standards to replicate and reproduce experiments and data \citep{schreiber2024trrraced}. Furthermore, the data collected following such protocols also enhances the comparability of the data, as participants of independent cohorts have been subject to the same stress-eliciting procedure.

We consider the following studies in more detail as they have made their data publicly available  and are lab-based. 

The Stress-Predict dataset presented by \cite{iqbal2022stress} introduces a novel dataset leveraging physiological data from wearable devices. This dataset comprises physiological responses of 35 subjects during approximately 60-minute experiments, in which the subjects faced different stress stimuli such as the Stroop Color Word Test, a semi-public speaking task (TSST), and a hyperventilation test. The authors emphasized that each experiment phase was labeled by pressing the button on the sensor device.

\cite{parent2020pass} present a comprehensive multi-modal physical activity and stress database, which has been made available to the research community. Using two different video games, the researchers differentiate between stress and the absence of stress. During the experiments, stress and its absence are moderated by varying physical activity levels, aiming to distinguish and recognize stress in the presence of physical activity or motion artifacts. The authors decided on video games instead of common stressors used in similar studies, such as the Stroop Color Word Test, to simulate a realistic task or stressor. However, the chosen stressful game involves navigating a seemingly abandoned, creepy asylum, evading harm and capture by its mutated inmates. The database significantly contributes to stress research by exploring the moderating effects of physical activity on stress levels. 

\cite{markova2019clas} introduced the CLAS dataset. This multi-modal dataset includes physiological responses (e.g., electrodermal activity, electrocardiography, and PPG) recorded during cognitive stimuli such as math problems, the Stroop Color Word Test, logic problems, and video clips. The authors point out that the protocol for collecting the CLAS dataset targets to evoke emotions in the quadrants of the well-established circumplex model \cite{russell1980circumplex}. The stimuli that evoke high cognitive loads during the study are similar to those found in many stress datasets. While the authors point out a detailed study protocol, labeling and annotation of the dataset are not discussed in detail.

Another publicly available dataset is provided by \cite{hongn2025wearable}. Similar to \citep{birjandtalab2016non}, this dataset aims to incorporate physiological responses during cognitive stress and physical activity to facilitate structural comparisons between cognitive and physical stress. The experimental protocol is carried out to evoke cognitive stress during several well-established stimuli (e.g., Stroop Color Word Test, mental arithmetic). Furthermore, the experimental protocol included anaerobic and aerobic exercises for a detailed analysis of physical activity. 

\cite{birjandtalab2016non} introduced the Non-EEG dataset to study the assessment of neurological status, including cognitive, emotional, and physical stress. The participants performed various tasks (e.g., arithmetic, watching video clips, and physical exercise) to induce the aforementioned neurological states. The researchers acknowledged that the dataset does not contain labels for the respective neurological states, highlighting that labels are not always given in real-life scenarios. 

The WESAD dataset was introduced by \cite{schmidt2018introducing} to study different affective states. The dataset was recorded in a controlled lab setting and bridges cognitive and socio-evaluative stress with an amusement state. The participants performed various tasks to induce stress or amusement (e.g., arithmetic, public speaking, watching video clips). Furthermore, the dataset includes informative metadata, self-reports of participants, and labels (i.e., events during the experiment). The different conditions served as the gold standard for stress assessment rather than the self-reports, although these are also included.

To our knowledge, none of the datasets (publicly available or not) employ a standardized experiment protocol. Most studies collecting datasets use similar sets of stimuli (e.g., TSST, Cold-Pressor) and experiment structures. However, most of the time, the study set-ups differ slightly, impeding the comparison between studies. \cite{schreiber2024trrraced} propose a framework to conduct stress-related experiments, which focuses on data collection and re-usability of such data.

\subsection{Automated Stress Detection}
\label{subsec:automated_stress_detection}
The task of automated stress detection in controlled settings has been addressed by many studies (e.g. \citep{birjandtalab2016non,hovsepian2015cstress,plarre2011continuous, schmidt2018introducing}. These studies have focused on lab-based induced stress, in which different stimuli are employed to elicit moderate stress in participants (e.g., physical stress, socio-evaluative stress, cognitive stress), while measuring signals that are emitted from the study subjects. These signals include brain waves, facial movements, speech, and physiological signals. Collected data can be employed for unsupervised or supervised machine learning algorithms to cluster or classify the different stimuli and stress events.

Research predominantly relies on binary classification to identify stress, yet this approach fails to distinguish between different types of stress. For instance, socio-evaluative stress induced through virtual interviews can be classified by analyzing voice signals, as shown in \citep{HafiyHilmy2021}, who reported an accuracy of 61\% . \cite{Arsalan2021} collected data during a public speaking task and demonstrate that physiological signals like electroencephalogram (EEG), galvanic skin response (GSR), and photoplethysmography (PPG) can enhance performance. They trained a Support Vector Machine (SVM) which yielded an accuracy of 96\%. Furthermore, since EEG is typically not worn continuously, the input data was reduced to only GSR and HR, which resulted in an accuracy of 86\%~\citep{Arsalan2021}.

Several studies approach binary classification utilizing the WESAD dataset which was introduced by \cite{schmidt2018introducing} to study different affective states. The dataset was recorded in a controlled lab setting and bridges cognitive and socio-evaluative stress with an amusement state. It includes informative meta-data, self-reports of participants and labels (i.e., events during the experiment). Schmidt et al. report an accuracy of 93\%, exploring several machine learning algorithms, with Linear Discriminant Analysis (LDA) yielding the most promising results \citep{schmidt2018introducing}. In addition,
Sah et al. have only used the electrodermal activity (EDA) signal to train a Convolutional Neural Network (CNN) model and achieved an accuracy of 93\% \citep{Sah2022}.

In contrast, \cite{koldijk2014swell} introduced the SWELL-KW dataset to study various stress types under cognitively demanding tasks. They collected multi-modal data such as physiological signals (e.g., HR, skin conductance), computer interactions (mouse, keystrokes), and self-reported stress. This dataset enables researchers to investigate how stress manifests during real-world work, supporting the development of stress recognition methods and adaptive user interfaces. Utilizing this dataset, \citep{Mortensen2023} classified between no stress, time pressure stress and interruption stress. A 1DCNN model was trained on only HRV data and achieved an accuracy of 99.9\%. It is claimed that the model is not overfitted, however, further validation with additional datasets is lacking~\citep{Mortensen2023}. 

Finally, two studies collect cognitive, emotional and physical stress, and relaxation states. \cite{birjandtalab2016non} introduced the Non-EEG dataset to study the assessment of neurological status. The participants perform various tasks (e.g., arithmetic, watching video clips, physical exercise) to induce the aforementioned stress types. However, the dataset does not contain labels for the respective neurological states which limits its usability. Thus, the authors employed the Gaussian Mixture Model (GMM), an unsupervised clustering method, and achieved an accuracy of 85\% \citep{birjandtalab2016non}. \cite{Moridani2020} worked with the same stress types, but only approach binary classification between relaxation and cognitive stress and between relaxation and emotional stress. Best performance is reported using a CNN model with 98\% and 94\%, respectively~\citep{Moridani2020}. 

In conclusion, it is evident that the majority of studies rely on a limited number of publicly available datasets, which are constrained in both features and classes. Therefore, most often binary classification is approached without distinguishing between different types of stress. Moreover, usually single input parameters are used, but incorporating additional vital parameters can provide deeper insights. While most studies demonstrate good performance with accuracy rates ranging from 85\% to 95\%, the field lacks validation of these results and generalization testing of models employed. In particular, validation with external datasets is impeded due to the diversity in the datasets and the varying conditions, under which they are collected. In addition, the combination of multiple datasets is restricted, which could increase the dataset size.
To our knowledge, none of the datasets employ a standardized experiment protocol. Most studies collecting datasets use similar sets of stimuli (e.g. Trier-Social-Stress-Test (TSST), Cold-Pressor) and experiment structures. However, most of the times, the study set-ups differ slightly, impeding the comparison between studies. \cite{schreiber2024trrraced} propose a framework to conduct stress related experiments, which focuses on data collection and re-usability of such. We argue that standardization allows for enhanced comparison and validation of research results by other researchers.

\section{Data and Methods}
\label{sec:data_and_methods}
The experimental setup and data collection are described in detail in the following. The study was approved by the local ethics committee.

\subsection{Anonymization}
\label{subsec:anonymization}
Each participant is assigned a randomly generated unique ID. The age of each participant is shown as an age range rather than an integer. The session dates and event marks during the protocols have been modified by a random time-deltas. Time samples have been shifted consistently across all records to maintain signal alignment.

\subsection{Participants}
\label{subsec:participants}
21 volunteers (18 male and 3 female with mean age of 24,1 and std of 3,97) participated in the data collection experiment. Upon registering for the experiment, the participants were sent an email containing general information and exclusion criteria. In this email, potential candidates were asked to wear sports clothing and refrain from drinking coffee and smoking one hour prior to the experiment. Before the start of the experiment, we double-checked the health status of the participants, informed them about the experiment, the data usage and storage policies, and asked for consent.

\subsection{Sensor Setup}
\label{subsec:sensor_setup}
Physiological data was collected using the Corsano Cardiowatch 287-2B System to collect physiological data. This wearable system is a Remote-Patient Monitoring System that consists of a monitoring device worn on the wrist. The bracelet is intended to monitor vital sign data continuously. The bracelet monitor offers flexibility to set data collection intervals by minute, second, or 25Hz, 32Hz, and 128Hz. The CardioWatch system can measure heart rate, RR intervals, breathing rate, A-Fib detection, SpO2, core body temperature, ECG, cuffless non-invasive blood pressure, activity, and sleep. The cardio watch 278-2 includes state-of-the-art PPG sensors with two diodes, green/red/IR LEDs, a one-read ECG, and a GSR-EDA sensor. The device is EU CE MDR medically certified, and FDA cleared\footnote{\url{https://corsano.com/products/bracelet-2/}}.

The sensors were placed on the weak hand-side's wrist to avoid additional motion artifacts while filling out the self-reports and preparing for the tasks. The experiments were not started with, until we received a signal from the Corsano wristband.

\subsection{Annotation Interface and Obtaining Ground Truth}
\label{subsec:annotation}
Obtaining the ground truth or a gold standard is a difficult endeavor with regards to determining the level of and stimulation of emotion, affect and stress. Several methods are used to assess stress levels, such as self-reports \citep{cohen1994perceived}, continuous labeling \citep{sharma2019dataset}, using the stimuli as a label or gold standard (e.g., \citep{schmidt2018introducing, hovsepian2015cstress}), or biological markers \citep{kirschbaum1993trier}. While the former two allow to deduce the level of involvement, self-reports may be criticized because they rely on self-reported data \citep{mishra2020continuous}, thus subjective judgment. While typical self-reports only allow for individual assessments at specific moments of time, new methods have been proposed to introduce a continuous nature into the self-assessments. For example, \cite{sharma2019dataset} presents a joystick based self assessment, allowing subjects to continuously indicate intensity levels of valence and arousal on two different axes by moving around the joystick. This set-up, however, depends on the task or stimuli and the experimental environment. For data annotation and obtaining ground truth, we follow the proposition by \citep{schmidt2018introducing}, which entails the collection of self-reports and using the context information and \textit{type-of-stress-being-induced} to label the data. The self-assessments may be considered as complementary information revealing the degree of affective engagement to be used in the personalization of user specific models.

A custom \textit{data annotation interface} was developed to detail the intricacies of each experiment. Physiological responses are sampled frequently, such that details about the underlying experiment process are required to analyze changes and trends in the data. The annotation interface helped us record meta data about start and end of different phases, locations and body positions (e.g., sitting, standing, walking). The user could also add relevant comments with timestamps. Such meta data helps to account for uncertainties that researchers might face during the analysis of such complex data. The \textit{annotation data} contains the labels which can be deducted from the stress events. This provides further contextualization for other researchers to understand the data, and specifically what happened during the experiments.

\subsection{Experimental Protocol}
\label{subsec:experimental_protocol}
This study aimed to elicit three different types of stress (cognitive, socio-evaluative, and physical) in participants. We record a neutral baseline before the start of the experiment, resting phases after the stimuli, as well as a recovery phase after the experiment. We propose our study protocol on the basis of the TRRRACED framework \citep{schreiber2024trrraced} to follow the paradigm of collecting and publishing data in a standardized format.

\begin{figure}
\centering
\includegraphics[width=0.8\columnwidth]{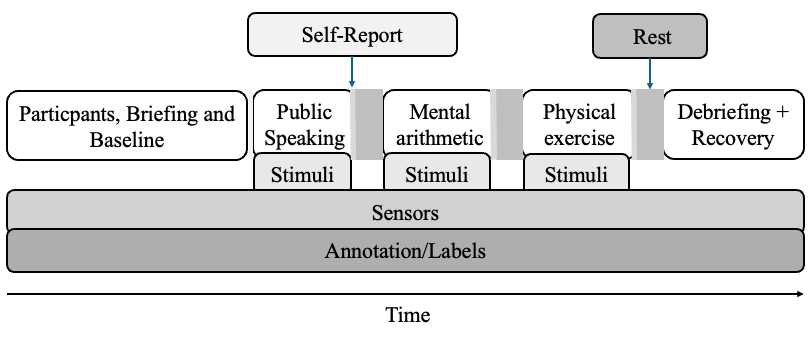}
\caption{Experimental Setup following \cite{schreiber2024trrraced}} \label{fig1}
\end{figure}

\textbf{Neutral}. The neutral baseline recording is a phase prior to the stress stimuli, aiming to induce the participants into a neutral psychological state. For this purpose, participants were seated on a comfortable chair and listened to calming sounds on noise-canceling headphones.

\textbf{Socio-evaluative}. The public speaking task elicits a socio-evaluative stress response based on the well-standardized Trier-Social-Stress-Test (TSST). During this task, participants are given 5 minutes to prepare a 5-minute speech about their skills, strength and weaknesses obtained during their studies to mimic a job interview \citep{kirschbaum1993trier}. Usually, this task is done in front of a jury, which supposedly judges the participant based on his performance. We conducted the public speaking task under the pretense that the performance would be later evaluated by psychologists. Thus, a camera and a microphone were positioned in the experiment room, without using them, to increase the socio-evaluative pressure, as is done in the original TSST \citep{kirschbaum1993trier}.

\textbf{Cognitive}. The mental arithmetic task aims to yield a cognitive stress response and is also based on the TSST \citep{kirschbaum1993trier}. During this task, participants were given the task of counting down from an arbitrary 4-digit number (e.g., 2037), deducting 13 until they are close to or reach zero. Every time the participants make a mistake, a noisy signal appeared, and they have to start all over again with a new number to subtract from a new arbitrary 4-digit number. The mistake is noted on the data annotation interface. In this way, we are trying to simulate a situation during which participants undergo cognitive stress. This task was also performed in front of the camera and the microphone, without any recordings being made.

\textbf{Physical}. The physical task aims to induce a physical stress response in the participants. It has been shown that physical activity correlates to cortisol production and secretion, and thus, it is highly related to stress in humans \citep{bozovic2013salivary}. The physical task in our study includes a warm-up followed by an incremental increase in speed on a treadmill. During this physical intervention, participants run at three different paces for 2 minutes each and 6 minutes in total. The procedure is based on the physical intervention used in \citep{birjandtalab2016non}, however, we also included a faster pace that would exceed fast walking. Changes in pace are noted within the data annotation interface.

\textbf{Self-Reports}.
Following the different stress stimuli and the neutral baseline, the participants were asked to rate how they felt by filling out a questionnaire. The self-assessment manikins questionnaire was used, as was done in similar experiments \citep{schmidt2018introducing}. The subjects rated their (emotional) state on three different 9-point-likert scales (valence, arousal, dominance) \citep{bradley1994measuring}. Usually, studies only consider the items regarding valence and arousal, as these have been shown to be associated with stress. In our experimental set-up, the dominance scale was also included. Perceived control might also affect the stress level of individuals, since helplessness has been linked to emotional deficits and an increased stress response \citep{maier1976learned}. The SAM can be viewed as the complementary questionnaire to generate data points for the circumplex model discussed earlier.

\textbf{Recovery}. During the recovery phase, participants were seated on a comfortable chair while listening to calming music for 5 minutes.

\subsection{Classification Algorithms}
\label{subsex:classification}
To evaluate the feasibility of automated stress recognition through commercial wearables, we used the preprocessed data provided by the Corsano platform. In this regard, six different modalities are combined into one dataset: heart rate, rr-interval, respiration, skin conductance, temperature, and accelerometer. The details of the pre-processing steps and classification tasks are listed below.

The pre-processing and analysis are carried out in Python 3.11.5 in Jupyter Notebook. The classification algorithms were imported from the scikit and XGBoost library.

The first step in the pre-processing pipeline consists of aligning the sampling frequencies of the different features. To avoid sparsity when concatenating the data, each feature is resampled to have a sampling frequency of 1/30 seconds (0.33Hz). After resampling, the different modalities are combined into one dataset for each subject. Next, the annotation files were used to annotate the physiological data of each participant, thus obtaining the proper target labels used for the supervised learning tasks. Following \cite{sharma2019dataset}, the physiological values of each subject are normalized to account for inter-personal differences.

\begin{figure}[htbp]
  \centering
  \includegraphics[width=0.9\linewidth]{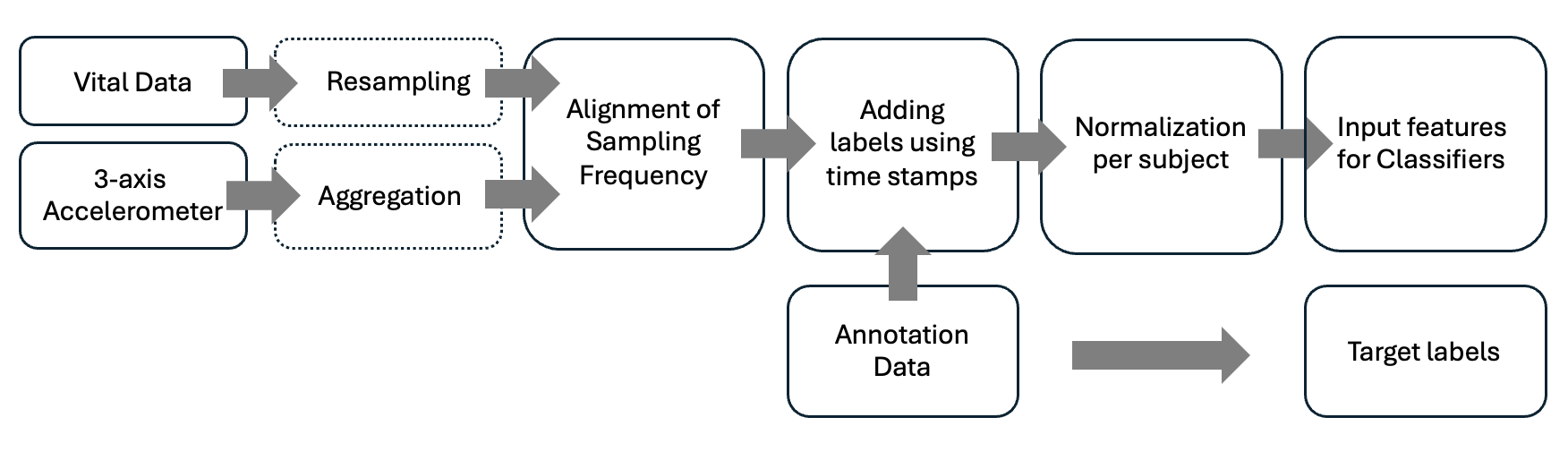}
    \caption{Pre-processing Pipeline}
\end{figure}

The features related to rr-intervals, as well as respiration rate, were removed because they contain a lot of missing data. Exploratory analysis showed that as soon as participants started moving around, the processed data files provided by Corsano contained a lot of missing data across participants. We speculate that motion artifacts interfere with the processing done by the Corsano system. 

The features and target labels described above serve as input for the classification algorithms. We evaluated the performance of six supervised classification algorithms for a binary and a three-class classification task. Several classifiers were compared: random forest (RF), linear discriminant analysis (LDA), k-Nearest-Neighbour (kNN), and Decision Tree (DT), as employed by \cite{schmidt2018introducing}. In addition, we also used Logistic Regression (LR) and XGBoost (XGB). Grid search with leave-one-subject-out cross-validation was performed for each classifier. The results reported here are the averaged accuracies of the different test splits. Hence, the results indicate how a global model would generalize and perform on data from previously unseen subjects. 

We evaluated our model using accuracy and F1-score. Accuracy measures the proportion of correctly classified instances among all samples. The F1-score, which is the harmonic mean of precision (a measure of result reliability for a given class) and recall (an indicator of completeness) was calculated by first determining precision and recall for each class and then averaging them. The F1-score is especially useful for unbalanced classification tasks, as is the case with the data at hand—where different conditions were recorded for varying durations during the study protocol.

\section{The VitaStress Dataset}
\label{sec:vitastress}
To enhance the usability of the dataset and provide other researchers with a detailed overview of the data and dataset structure, this section provides insight into the structure of the dataset repository and details the descriptive statistics of the dataset. The VitaStress Dataset is available at \url{https://github.com/paulvincenz/VitaStress}.

\subsection{Data and Summary Statistics}
\label{subsec:data_and_summary_stats}
There are two types of data exported from the Corsano portal, the vital parameters that are processed data provided by the CardioWatch System, and the raw data that come directly from the sensors. All files exported and available in the dataset are csv files. For more information on the raw and processed corsano files, we refer the reader to the corsano website \footnote{\url{https://developer.corsano.com/android/category/data-models}}.

The data recordings for the 21 subjects comprise a total of 1072 minutes worth of physiological signals and annotations. The total time elapsed during stimuli, including the neutral baseline recording, adds up to 542.5 minutes of physiological signals.

\begin{figure}
\centering
\includegraphics[width=\columnwidth]{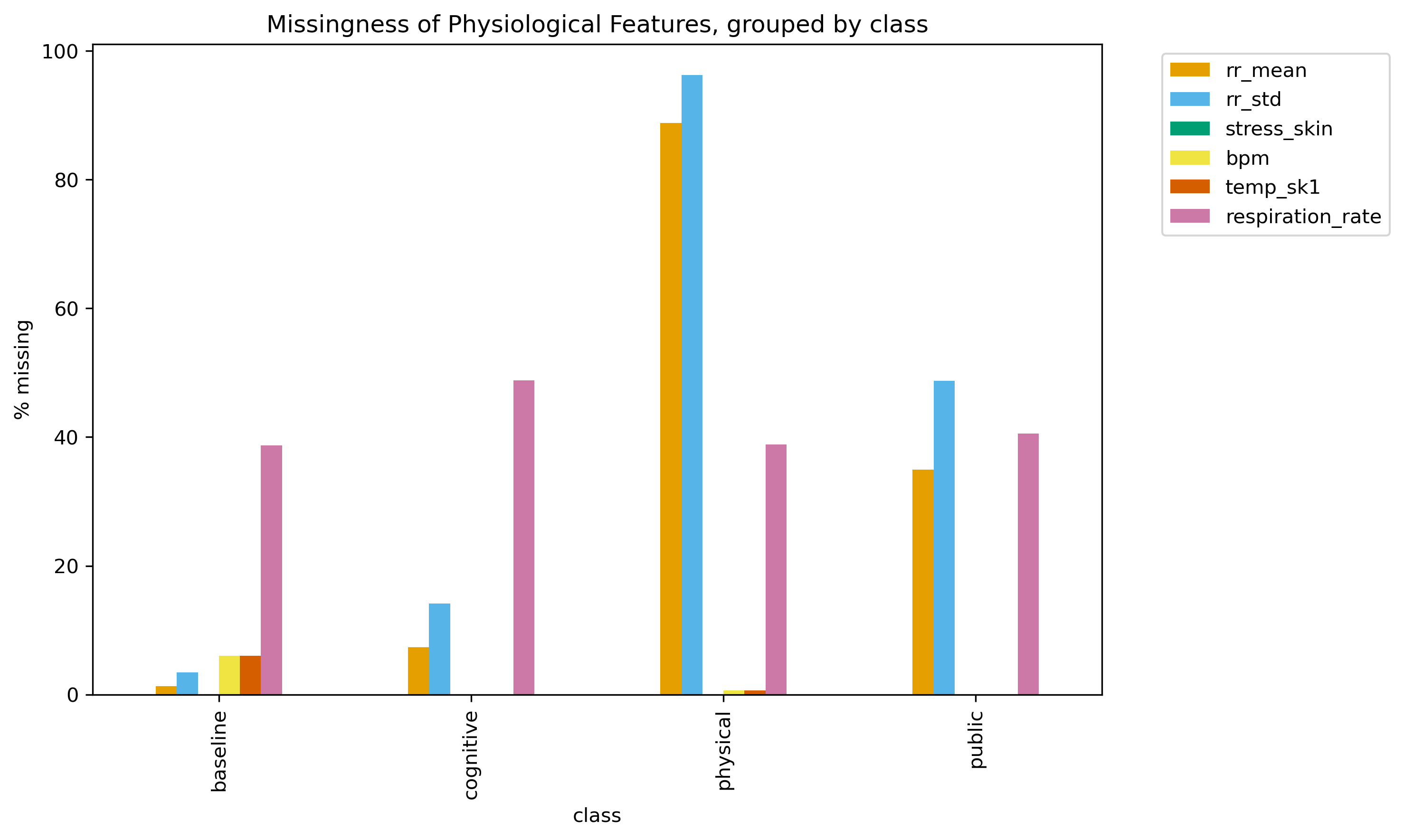}
\caption{The bar chart shows the percentage of missing values for six physiological variables - mean rr-interval (rr\_mean), rr-interval standard deviation (rr\_std), skin temperature (temp\_sk1), electrodermal activity (stress\_skin) and respiration rate (respiration\_rate) - across four conditions (baseline, cognitive load, physical exertion, and public speaking)}
\label{missingness-physio}
\end{figure}

The processed physiological signals contain high percentages of missing data for the features related to RR-intervals and respiration rate (see Figure \ref{missingness-physio}). The heart-rate variability (HRV) metrics (rr\_mean and rr\_std) exhibit substantial data loss, especially during physical exertion (>85\% missing), whereas heart rate and skin temperature recordings are nearly complete in all conditions (<5\% missing). The electrodermal activity shows minimal missing values (<5\%), and the aggregated respiration\_rate feature has moderate gaps (~40–50\%) across all classes. While the missingness of the respiration\_rate is relatively stable over the different class labels, the missingness of the rr-interval features peaks during the physical and socio-evaluative stimuli. This indicates that the preprocessing system cannot handle increased physical activity or motion artifacts, leading to a decrease in feature quality, and thus, the system outputs NA values.

\subsubsection{Feature Validation}
\label{subsubsec:feature_validation}
\paragraph{Accelerometry} Under baseline conditions, the mean accelerometer X and mean accelerometer Y distributions are centered near zero with moderate variance, reflecting minimal net movement. In the cognitive condition, the acceleration along the X and Y axes exhibit markedly reduced spread around zero, indicating restricted motion during focused mental tasks. Physical exertion induces a pronounced negative shift in accX\_mean (median < 0) and the greatest variance in both X- and Y-axis accelerations, consistent with vigorous, multidirectional movement. The public condition yields intermediate acceleration profiles: moderate variance around zero in acceleration along the X axis and a slight positive shift in acceleration along the Y axis.

The distributions of the acceleration along the Z axis show that baseline recordings are skewed positively, reflecting the influence of gravity on the sensor, whereas cognitive trials are narrowly distributed close to zero. Physical and public conditions also show positive medians but with greater variability under physical load.

\paragraph{Electrodermal Activity} Skin‐conductance responses are highest in the cognitive condition, with both the greatest median and widest distribution, suggesting elevated sympathetic arousal during mental workload. Public speaking elicits a moderate increase in skin conductance relative to baseline, whereas physical exertion does not substantially elevate stress\_skin beyond rest, possibly due to sweat‐induced measurement artifacts.

\paragraph{Cardiac Response} Heart rate increases progressively from baseline through cognitive and public conditions, reaching its maximum median and variance during physical exertion. This gradient aligns with metabolic demand: minimal cardiovascular activation at rest, moderate mental and social stress increases, and peak rates under physical activity.

\paragraph{Skin Temperature} Skin temperature is highest during physical activity, consistent with heat generation and peripheral vasodilation, and lowest during cognitive tasks, potentially reflecting radiation cooling during periods of low muscular activity. The physiological responses to the socio-evaluative stressor again occupy an intermediate position, and baseline temperatures lie between the means of the socio-evaluative and cognitive conditions.

\begin{figure}
\centering
\includegraphics[width=\columnwidth]{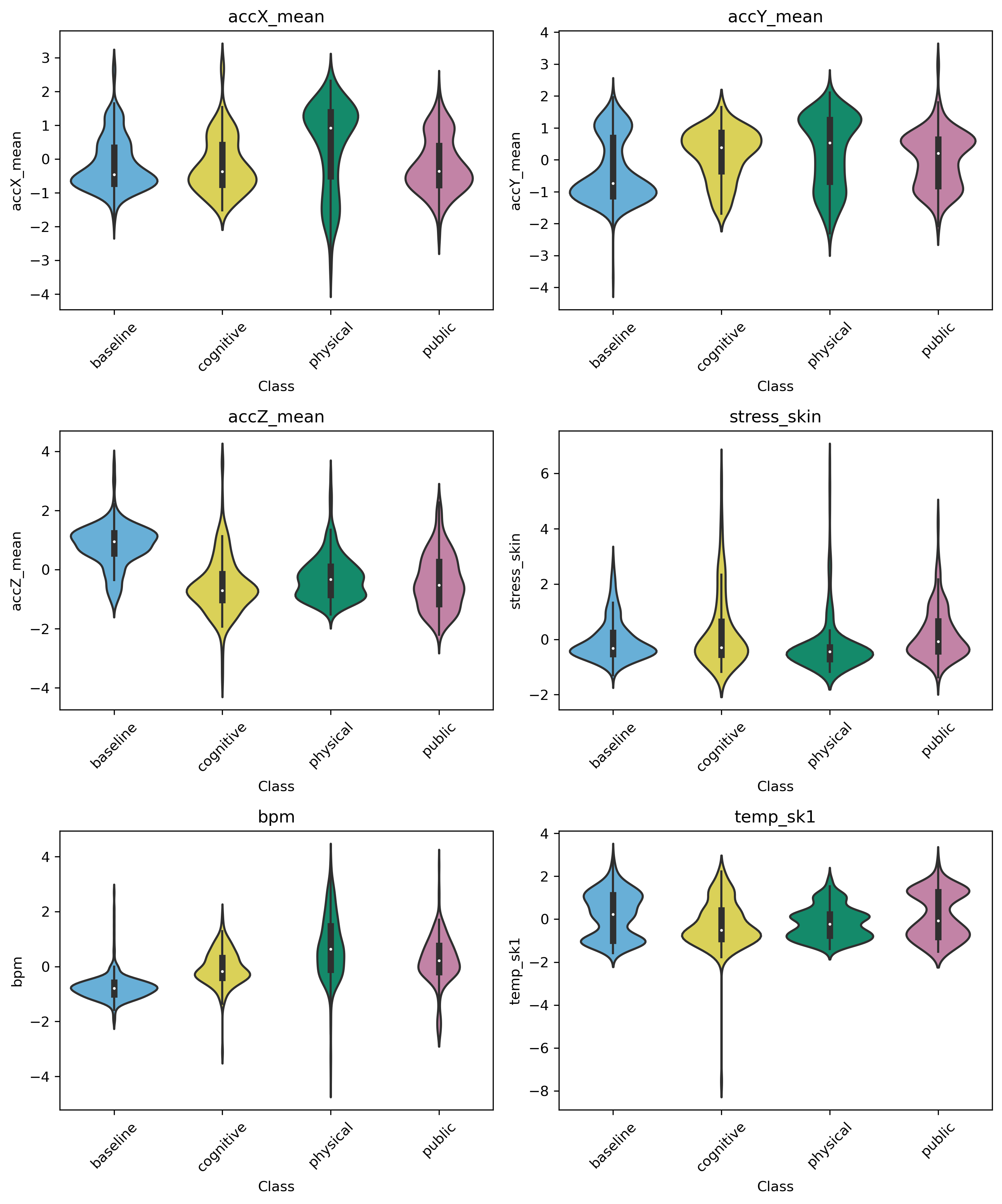}
\caption{“Violin” plots of the distribution of the selected features across the different phases of the experiments. The box plots embedded in each violin plot show the Interquartile Range (IQR) for each considered variable, while a yellow diamond marks the mean of the distribution.} \label{violinplots_features}
\end{figure}

\subsubsection{Principal Component Analysis}
\label{subsubsec:principal_component_analysis}
\begin{figure}
\centering
\includegraphics[width=0.9\columnwidth]{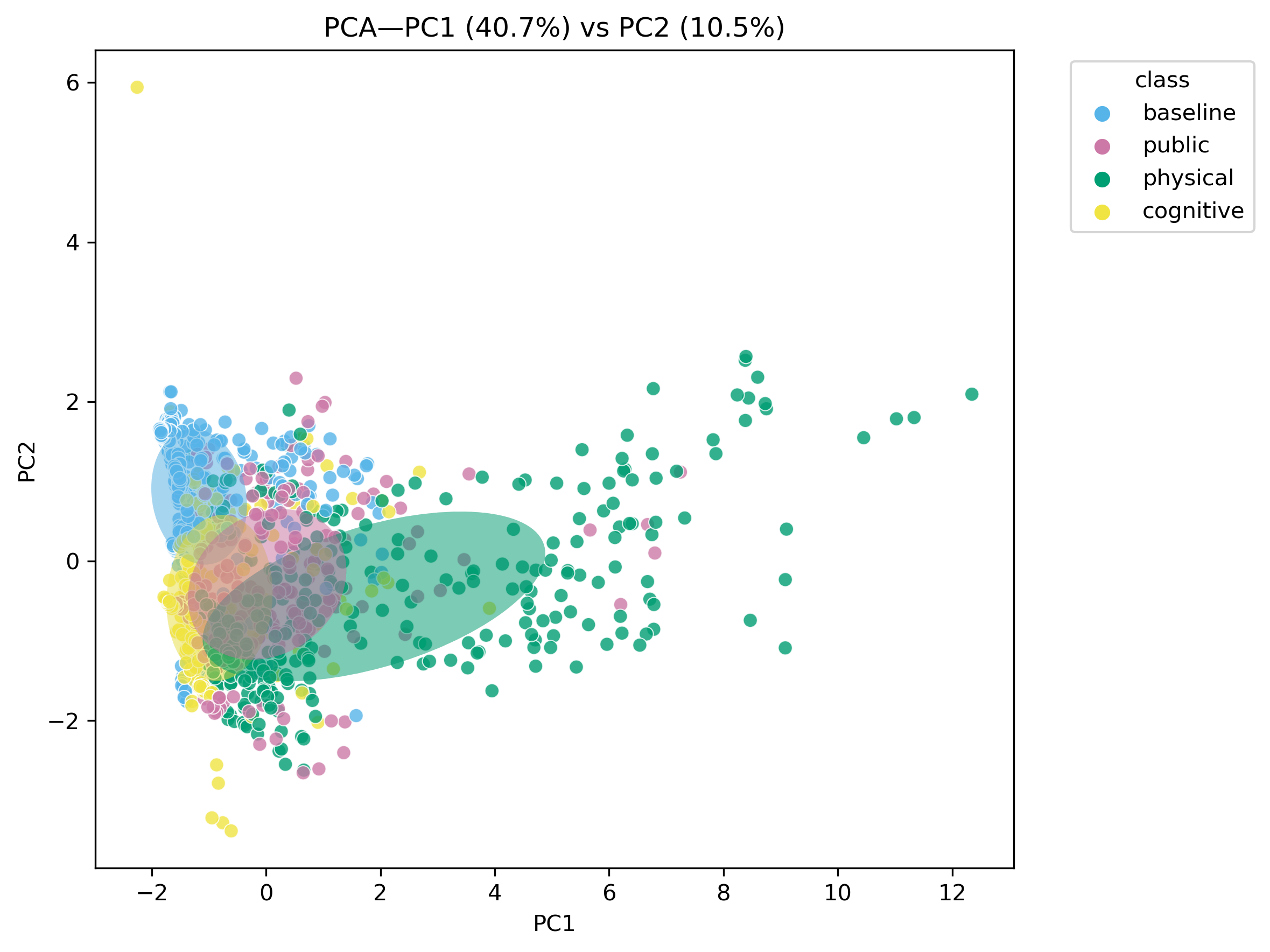}
\caption{Principal component analysis of kinematic and physiological features collected under four conditions.} \label{pca}
\end{figure}

We applied principal component analysis (PCA) (see Figure \ref{pca}) to the set of selected z-scored variables—namely the mean, standard deviation, and variance of acceleration along the X, Y, and Z axes, electrodermal activity, heart rate (bpm), and skin temperature—collected under four experimental conditions (baseline, cognitive, public, and physical). The first two principal components (PCs) accounted for 40.7\% and 10.5\% of the total variance, respectively, for a cumulative explained variance of 51.2\%.

PC1 loads strongly and positively on all accelerometer variability measures (accX\_std: 0.431, accX\_var: 0.402, accY\_std: 0.419, accY\_var: 0.398, accZ\_std: 0.372, accZ\_var: 0.344) and on heart rate (bpm: 0.234), with near-zero contributions from the mean accelerations and skin-conductance. Thus, PC1 captures a composite “activity intensity” signal that combines gross motor variability with cardiovascular activation. In the PC1–PC2 space, physical exertion trials project far to the right (high PC1), whereas baseline, cognitive, and public conditions cluster near zero or on the left, reflecting lower movement and heart-rate demands.

PC2 contrasts mean vertical acceleration (accZ\_mean: +0.680) and, to a lesser extent, electrodermal arousal (+0.126) against lateral acceleration bias (accY\_mean: –0.537), heart rate (bpm: –0.387), and skin temperature (–0.155). High PC2 scores thus indicate a predominantly static, upright phenotype with low thermal and cardiac load, whereas low PC2 scores reflect lateral sway combined with elevated cardiac and thermal responses. Accordingly, baseline recordings occupy the upper PC2 region (static posture, low stress), cognitive trials lie at the lower extreme (minimal vertical motion but high skin conductance), and public speaking and physical exertion span intermediate PC2 values.

\subsubsection{Self-assessments}
\label{subsubsec:self_assessments}
\begin{figure}
\centering
\includegraphics[width=0.8\columnwidth]{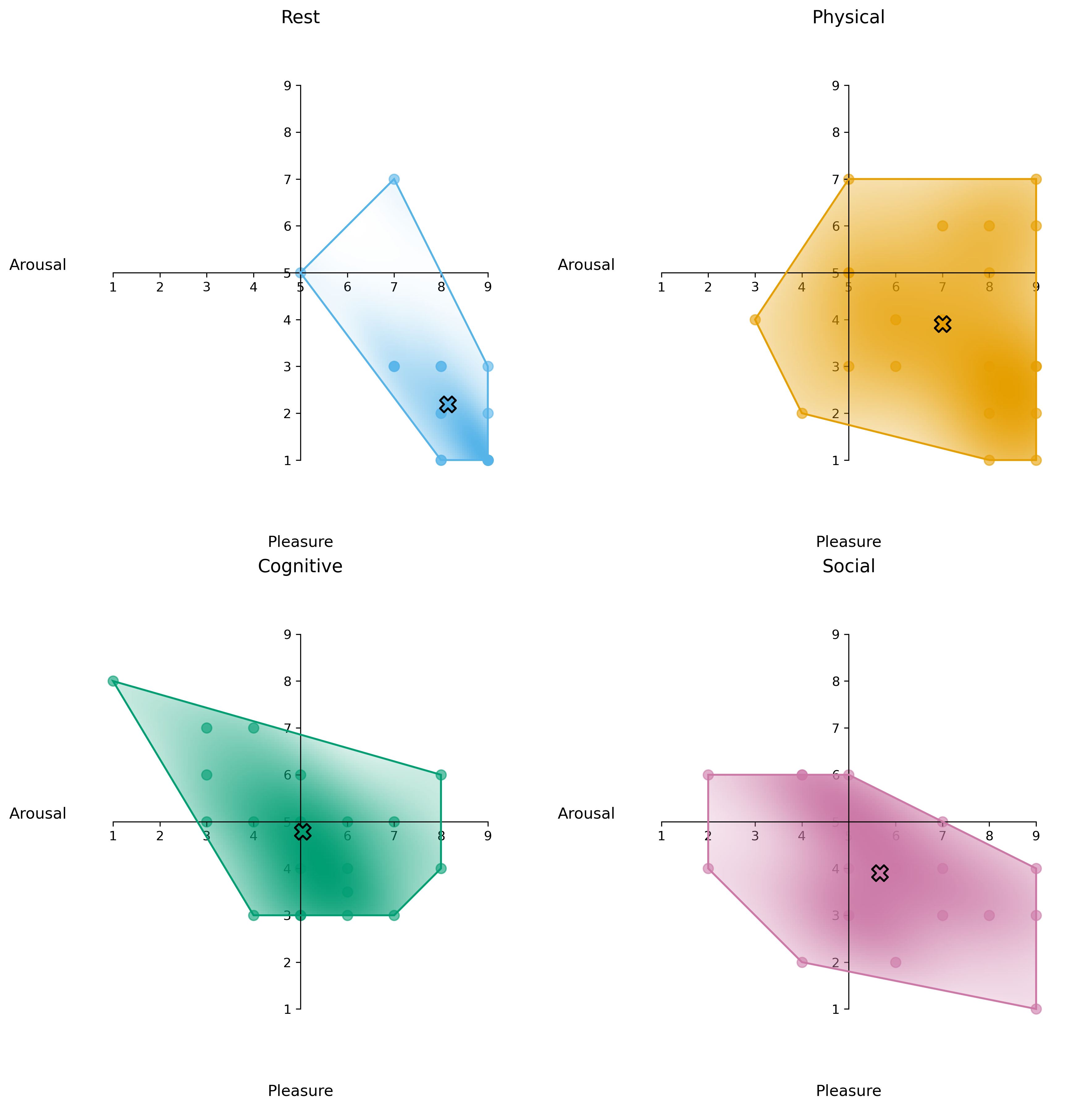}
\caption{The subplots in this figure show the subjective ratings for the different phases of the experiment. The shaded areas in the hull convex polygons indicate the density of similar ratings, the lesser shaded areas show that less people provided similar assessments.} \label{circumplex_results}
\end{figure}

Figure \ref{circumplex_results} shows the individual mappings of the subjects, the centroids of the self-assessments, and the distribution of the self-assessments as convex hull polygons, grouped by stress-stimuli. While none of the centroids are situated outside the low arousal and high valence quadrant, which is associated with a relaxed affective state, each stimulus exhibits its characteristics and tendencies. The self-reports show that the neutral baseline recording at the beginning of each experiment yielded mild happy and relaxed states, with most data points at the plot's border at the lower right-hand side. The self-reports for the physical activity show that two participants felt rather sad and fatigued. In contrast, most subjects felt relaxed, with a tendency towards higher arousal and, hence, a happy affective state. The self-reports provided for the cognitive stress stimuli depict a diverse participant assessment. While the centroid is located around 5.05 on the valence-axis and 4.78 on the arousal-axis, with the majority of subjects rating their affective state similarly, the polygon in c) clearly shows a tendency towards the high arousal/low valence quadrant usually associated with an angry affective state but also stress, as shown in Figure \ref{circumplex}. The self-assessments following the socio-evaluative stressor exhibit tendencies towards the high arousal/low valence quadrant associated with stress and the quadrant associated with sadness. The centroid and the majority of the subjects provided a relatively relaxed assessment, as shown in d) of Figure \ref{circumplex_results}. In addition to this brief analysis of the self-assessments on the standard valence-arousal model, a third orthogonal axis could be added to account for the assessment on the \textit{dominance} item. 

\subsection{Data Files and Structures}
\label{subsec:data_files_and_structures}
The VitaStress repository contains the physiological data, self-reports, as well as the annotation data of all 21 subjects. The repository contains folders for each subject, which contain the respective recordings. The \textit{id\_activity.csv} files contain the processed data, whereas the other files contain the raw data from the sensors. Moreover, each subject folder contains the \textit{id\_annotation.csv} and the \textit{id\_selfreport\_.csv} files.

The annotation files contain meta-data about the experiments. We suggest that researchers and others use this meta-data to label the physiological data. As explained in section \ref{subsec:annotation}, this meta-data includes information such as start and end of main stimuli, locations, and body positions. The annotation files are structured as follows:
\begin{itemize}
    \item \textbf{\textit{Column 1}} Time-stamps. 
    \item \textbf{\textit{Column 2}} Start/end of stimuli, locations, body positions, and comments.
\end{itemize}

The self-report files contain the self assessments provided by each subject during the experiment. As mentioned above, the values range from 0 to 9, where 0 corresponds to negative, 5 to neutral, and 9 to positive of the given item.

\begin{itemize}
    \item \textbf{\textit{Column 1-3.}} These columns indicate the arousal, valence, and dominance values (corresponding order) given by a subject. 
    \item \textbf{\textit{Row 1-4.}} These rows indicate self-assessed values across the different conditions: neutral baseline, physical activity, cognitive stimuli, socio-evaluative stimuli (corresponding order). 
\end{itemize}

\section{Results}
\label{sec:results}
Based on the different types of stress elicited during the experiments, we distinguish between two different classification tasks: three class classification (neutral baseline, physical, cognitive 
 \& social stress) and a binary classification problem (neutral baseline and stress). In our analysis, the neutral baseline class is representative of the absence of stress or \textit{neutral} affect. Regarding the binary classification, we map physical, cognitive and socio-evaluative stress to a class called \textit{stress}. For each classification task we evaluate the performance of the classifiers using only physiological responses, only motion, and physiological and motion together as input features. The results of the three class classification task are shown in Table \ref{tab1} and Table \ref{tab2}, and the results of the binary classification task are presented in Table \ref{tab2} and Table \ref{tab4}.

Regarding the ternary classification task, kNN yields the overall best performance with 79\% in accuracy and 74\% F1-score. Following kNN, XGBoost shows the next best performances with 77\% accuracy and 0.74 in F1-score. As seen in Tables \ref{tab1} and \ref{tab2}, different classifiers achieve the best results depending on the input features. Notably, kNN performs best with just motion features as input. This indicates that the vital data are not informative for the algorithm, or that trends in the vital data are too subtle for the classifier to utilize.

\begin{table}[ht]
\caption{Binary Classification - F1-Score per Classifier.}\label{tab1}
\centering
\small
\fontsize{7}{10}
\resizebox{\columnwidth}{!}{%
\begin{tabular}{lclllllllllll}
%\hline
\multicolumn{1}{c}{} &
  \multicolumn{12}{c}{Accuracy} \\ \cline{1-13} 
 &
  \multicolumn{2}{c}{LR} &
  \multicolumn{2}{c}{RF} &
  \multicolumn{2}{c}{DT} &
  \multicolumn{2}{c}{LDA} &
  \multicolumn{2}{c}{XGB} &
  \multicolumn{2}{c}{kNN} \\ \hline
All &
  \multicolumn{2}{c}{$ \num{0.76914211154003} 
  \pm \num{0.16233418813594538}$} &
  \multicolumn{2}{c}{$ \num{0.7723476514545118} 
  \pm \num{0.19131933941731888}$} &
  \multicolumn{2}{c}{$ \num{0.6816646605893023} 
  \pm \num{0.21606221855709085}$} &
  \multicolumn{2}{c}{$ \num{0.7579714018301668} 
  \pm \num{0.14874161454120016}$} &
  \multicolumn{2}{c}{$ \textbf{\num{0.7727565970026592}}
  \pm \num{0.18308600165992062}$} &
  \multicolumn{2}{c}{$ \num{0.7628258988312734} 
  \pm \num{0.20503949481090145}$} \\
Physio &
  \multicolumn{2}{l}{$ \num{0.6313469164124149} 
  \pm \num{0.1472102341523598}$} &
  \multicolumn{2}{l}{$ \textbf{\num{0.6580359624389619}}
  \pm \num{0.15730012708255892}$} &
  \multicolumn{2}{l}{$ \num{0.628213437407098} 
  \pm \num{0.14223464121702686}$} &
  \multicolumn{2}{l}{$ \num{0.6452459218161574} 
  \pm \num{0.14748465644568914}$} &
  \multicolumn{2}{l}{$ \num{0.6505551823920199} 
  \pm \num{0.1477682157314895}$} &
  \multicolumn{2}{l}{$ \num{0.6270119459387891} 
  \pm \num{0.13985427354466096}$} \\
Motion &
  \multicolumn{2}{l}{$ \num{0.7263952038463507} 
  \pm \num{0.19322442537510873}$} &
  \multicolumn{2}{l}{$ \num{0.6873446880844112} 
  \pm \num{0.22627830250654327}$} &
  \multicolumn{2}{l}{$ \num{0.6860137887055384} 
  \pm \num{0.21247633928695042}$} &
  \multicolumn{2}{l}{$ \num{0.7303151202762456} 
  \pm \num{0.1721531566809755}$} &
  \multicolumn{2}{l}{$ \num{0.7488941735967531} 
  \pm \num{0.2077346957563024}$} &
  \multicolumn{2}{l}{$ \textbf{\num{0.7865950949593431}} 
  \pm \num{0.18250942536986167}$} \\ \hline
\end{tabular}%
}
\end{table}

\begin{table}[ht]
\caption{Ternary Classification - F1-Score per Classifier.}\label{tab2}
\centering
\fontsize{7}{10}
\resizebox{\columnwidth}{!}{%
\begin{tabular}{lclllllllllll}
%\hline
\multicolumn{1}{c}{} &
  \multicolumn{12}{c}{F1-Score} \\ \cline{1-13} 
 &
  \multicolumn{2}{c}{LR} &
  \multicolumn{2}{c}{RF} &
  \multicolumn{2}{c}{DT} &
  \multicolumn{2}{c}{LDA} &
  \multicolumn{2}{c}{XGB} &
  \multicolumn{2}{c}{kNN} \\ \hline
All &
  \multicolumn{2}{c}{$ \num{0.7422357197978868} 
  \pm \num{0.18532017136382214}$} &
  \multicolumn{2}{c}{$ \num{0.7372551993745484} 
  \pm \num{0.22957719231341794}$} &
  \multicolumn{2}{c}{$ \num{0.653885220493795} 
  \pm \num{0.241120739062607}$} &
  \multicolumn{2}{c}{$ \num{0.7338017244520086} 
  \pm \num{0.17265369321012822}$} &
  \multicolumn{2}{c}{$ \textbf{\num{0.7390684321983286}}
  \pm \num{0.2189933242111045}$} &
  \multicolumn{2}{c}{$ \num{0.736459227259848} 
  \pm \num{0.23207276496603238}$} \\
Physio &
  \multicolumn{2}{l}{$ \num{0.6008117791930259} 
  \pm \num{0.16065163504211968}$} &
  \multicolumn{2}{l}{$ \textbf{\num{0.6154268622549426}}
  \pm \num{0.1647668235951791}$} &
  \multicolumn{2}{l}{$ \num{0.5841261727989693} 
  \pm \num{0.1578589908719607}$} &
  \multicolumn{2}{l}{$ \num{0.6106714827580889} 
  \pm \num{0.15598764573978255}$} &
  \multicolumn{2}{l}{$ \num{0.6046929166684389} 
  \pm \num{0.15375367136078308}$} &
  \multicolumn{2}{l}{$ \num{0.5926076020083466} 
  \pm \num{0.1404468186497193}$} \\
Motion &
  \multicolumn{2}{l}{$ \num{0.7046040623066917} 
  \pm \num{0.21429210205938548}$} &
  \multicolumn{2}{l}{$ \num{0.6459891871837693} 
  \pm \num{0.25647603816686715}$} &
  \multicolumn{2}{l}{$ \num{0.6556095779790717} 
  \pm \num{0.23833866546662397}$} &
  \multicolumn{2}{l}{$ \num{0.7088217106969664} 
  \pm \num{0.19584072778501074}$} &
  \multicolumn{2}{l}{$ \num{0.7140373084757257} 
  \pm \num{0.2388768330109614}$} &
  \multicolumn{2}{l}{$ \textbf{\num{0.7449748598955319}} 
  \pm \num{0.22582428510596703}$} \\ \hline
\end{tabular}%
}
\end{table}

For the binary classification task, random forest achieves the highest accuracy with 89\% and and F1-score of 0.82. (see Tables \ref{tab3} and \ref{tab4}). The random forest algorithm achieves the highest accuracies with all input features and only motion features, however, XGBoost shows the best performance in terms of accuracy and F1-score when the vital data are the only input features. Comparing the confusion matrices for both classifiers, it becomes apparent that XGBoost wrongfully classified \textit{stress} as the \textit{neutral} class more often than random forest did.

\begin{table}[ht]
\caption{Binary Classification - Accuracy per Classifier.}\label{tab3}
\centering
\fontsize{7}{10}
\resizebox{\columnwidth}{!}{%
\begin{tabular}{lclllllllllll}
%\hline
\multicolumn{1}{c}{} &
  \multicolumn{12}{c}{Accuracy} \\ \cline{1-13} 
 &
  \multicolumn{2}{c}{LR} &
  \multicolumn{2}{c}{RF} &
  \multicolumn{2}{c}{DT} &
  \multicolumn{2}{c}{LDA} &
  \multicolumn{2}{c}{XGB} &
  \multicolumn{2}{c}{kNN} \\ \hline
All &
  \multicolumn{2}{c}{$ \num{0.8773404979053151} 
  \pm \num{0.12714486834858135}$} &
  \multicolumn{2}{c}{$ \textbf{\num{0.8869776265819551}} 
  \pm \num{0.10822577689470238}$} &
  \multicolumn{2}{c}{$ \num{0.8791043379189285} 
  \pm \num{0.1255014683877096}$} &
  \multicolumn{2}{c}{$ \num{0.8765088128426589} 
  \pm \num{0.10067826743568017}$} &
  \multicolumn{2}{c}{$ \num{0.8699718516154843}
  \pm \num{0.14159604188597386}$} &
  \multicolumn{2}{c}{$ \num{0.8677348477833213} 
  \pm \num{0.13450105675966786}$} \\
Physio &
  \multicolumn{2}{l}{$ \num{0.8212253554702753} 
  \pm \num{0.1296105688149102}$} &
  \multicolumn{2}{l}{$ \num{0.8348613367863635} 
  \pm \num{0.124641800837161}$} &
  \multicolumn{2}{l}{$ \num{0.8135227115885751} 
  \pm \num{0.14090836381009653}$} &
  \multicolumn{2}{l}{$ \num{0.6452459218161574} 
  \pm \num{0.13347553956205516}$} &
  \multicolumn{2}{l}{$ \textbf{\num{0.8448863906405296}} 
  \pm \num{0.11012664514532627}$} &
  \multicolumn{2}{l}{$ \num{0.78886842224786} 
  \pm \num{0.12130300256202936}$} \\
Motion &
  \multicolumn{2}{l}{$ \num{0.8773404979053151} 
  \pm \num{0.12714486834858135}$} &
  \multicolumn{2}{l}{$ \textbf{\num{0.8869776265819551}} 
  \pm \num{0.10822577689470238}$} &
  \multicolumn{2}{l}{$ \num{0.8791043379189285} 
  \pm \num{0.10067826743568017}$} &
  \multicolumn{2}{l}{$ \num{0.8765088128426589} 
  \pm \num{0.1255014683877096}$} &
  \multicolumn{2}{l}{$ \num{0.8699718516154843} 
  \pm \num{0.14159604188597386}$} &
  \multicolumn{2}{l}{$ \num{0.8677348477833213} 
  \pm \num{0.13450105675966786}$} \\ \hline
\end{tabular}%
}
\end{table}

\begin{table}[ht]
\caption{Ternary Classification - Accuracy per Classifier.}\label{tab4}
\centering
\fontsize{7}{10}
\resizebox{\columnwidth}{!}{%
\begin{tabular}{lclllllllllll}
%\hline
\multicolumn{1}{c}{} &
  \multicolumn{12}{c}{F1-Score} \\ \cline{1-13} 
 &
  \multicolumn{2}{c}{LR} &
  \multicolumn{2}{c}{RF} &
  \multicolumn{2}{c}{DT} &
  \multicolumn{2}{c}{LDA} &
  \multicolumn{2}{c}{XGB} &
  \multicolumn{2}{c}{kNN} \\ \hline
All &
  \multicolumn{2}{c}{$ \num{0.8424085144217381} 
  \pm \num{0.16494535950018213}$} &
  \multicolumn{2}{c}{$ \textbf{\num{0.8202946594378279}} 
  \pm \num{0.20238583065163973}$} &
  \multicolumn{2}{c}{$ \num{0.8403742103057139} 
  \pm \num{0.16326425438923867}$} &
  \multicolumn{2}{c}{$ \num{0.8243118038884276} 
  \pm \num{0.17355434333221845}$} &
  \multicolumn{2}{c}{$ \num{0.8136664441902468}
  \pm \num{0.20462072099722695}$} &
  \multicolumn{2}{c}{$ \num{0.8204944627126057} 
  \pm \num{0.19198415830308393}$} \\
Physio &
  \multicolumn{2}{l}{$ \num{0.7458766977352462} 
  \pm \num{0.2102284768862876}$} &
  \multicolumn{2}{l}{$ \num{0.7790969127804165} 
  \pm \num{0.18940675697651357}$} &
  \multicolumn{2}{l}{$ \num{0.7679003429759786} 
  \pm \num{0.19378118972809538}$} &
  \multicolumn{2}{l}{$ \num{0.6953306071502949} 
  \pm \num{0.2165747716803555}$} &
  \multicolumn{2}{l}{$ \textbf{\num{0.7870993592965237}} 
  \pm \num{0.1759273867032519}$} &
  \multicolumn{2}{l}{$ \num{0.7711226812027386} 
  \pm \num{0.16967750771683265}$} \\
Motion &
  \multicolumn{2}{l}{$ \num{0.8424085144217381} 
  \pm \num{0.16494535950018213}$} &
  \multicolumn{2}{l}{$ \textbf{\num{0.8202946594378279}} 
  \pm \num{0.20238583065163973}$} &
  \multicolumn{2}{l}{$ \num{0.8243118038884276} 
  \pm \num{0.17355434333221845}$} &
  \multicolumn{2}{l}{$ \num{0.8403742103057139} 
  \pm \num{0.16326425438923867}$} &
  \multicolumn{2}{l}{$ \num{0.8136664441902468} 
  \pm \num{0.20462072099722695}$} &
  \multicolumn{2}{l}{$ \num{0.8204944627126057} 
  \pm \num{0.19198415830308393}$} \\ \hline
\end{tabular}%
}
\end{table}

\begin{figure}[ht]
  \centering
  \begin{minipage}{0.45\textwidth}
    \centering
    \includegraphics[width=\linewidth]{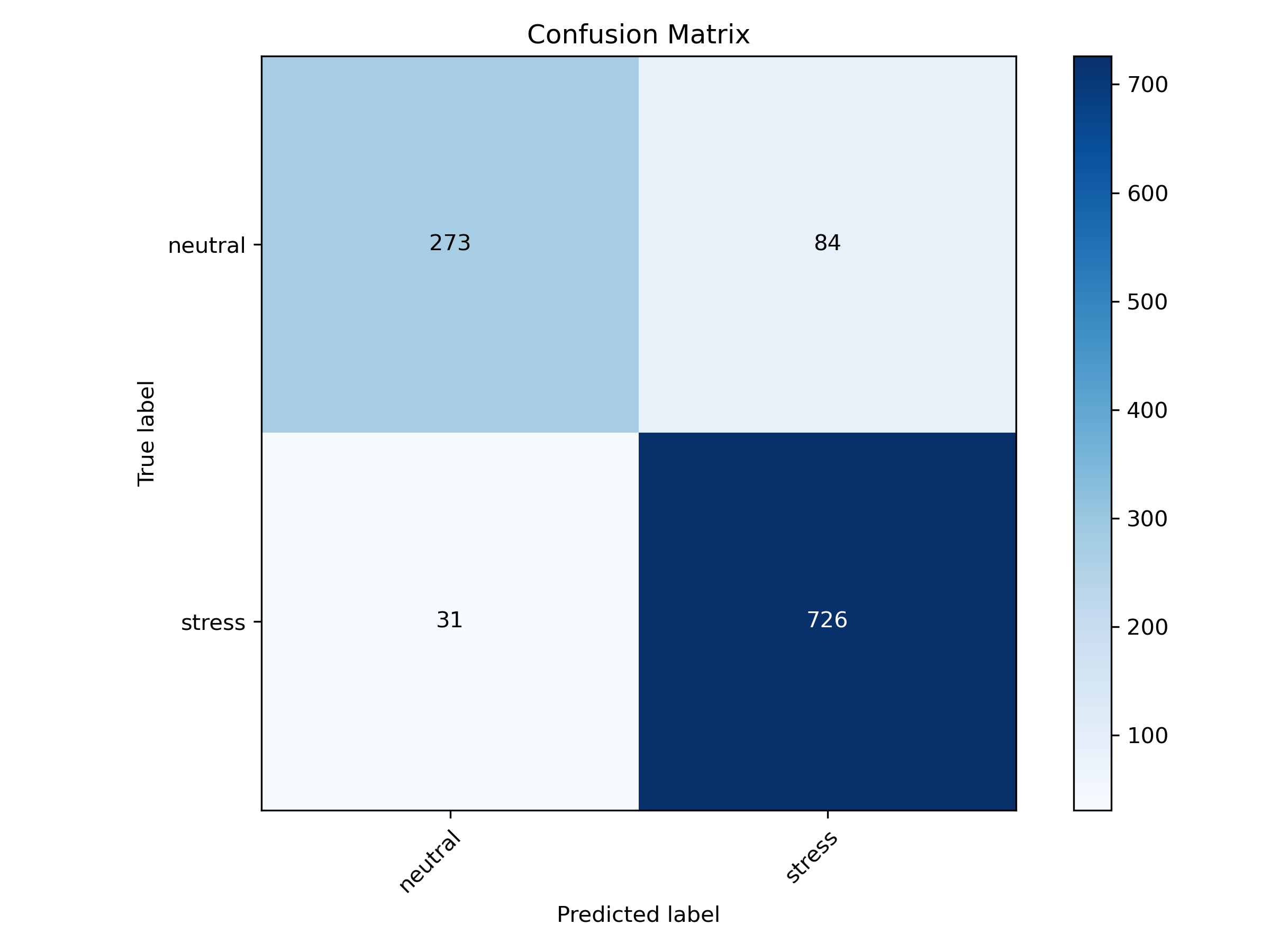}
    \caption{Confusion Matrix for kNN}
  \end{minipage}\hfill
  \begin{minipage}{0.45\textwidth}
    \centering
    \includegraphics[width=\linewidth]{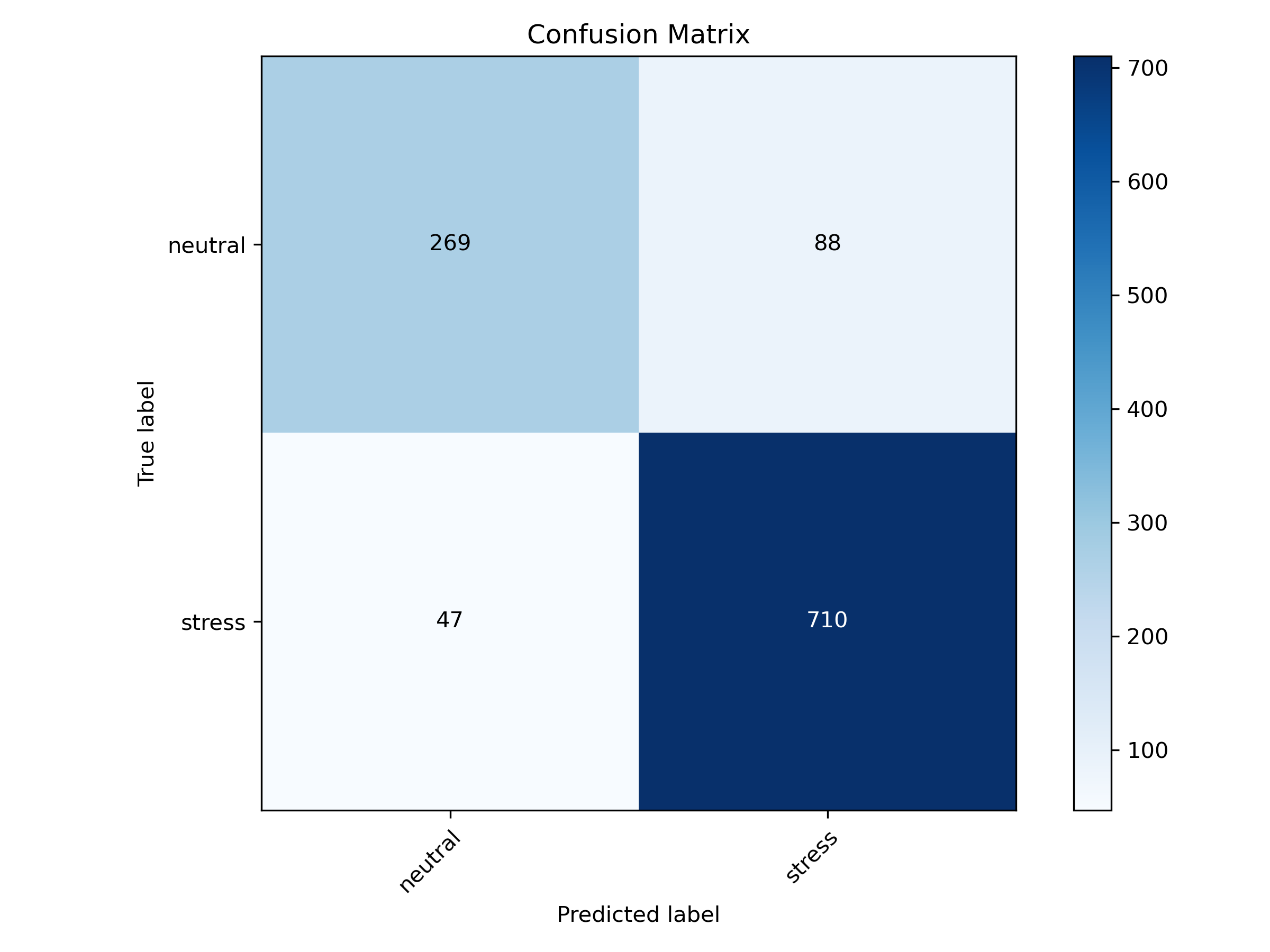}
    \caption{Confusion Matrix for XGBoost}
  \end{minipage}\hfill
\end{figure}

\begin{figure}[ht]
  \centering
  \begin{minipage}{0.3\textwidth}
    \centering
    \includegraphics[width=\linewidth]{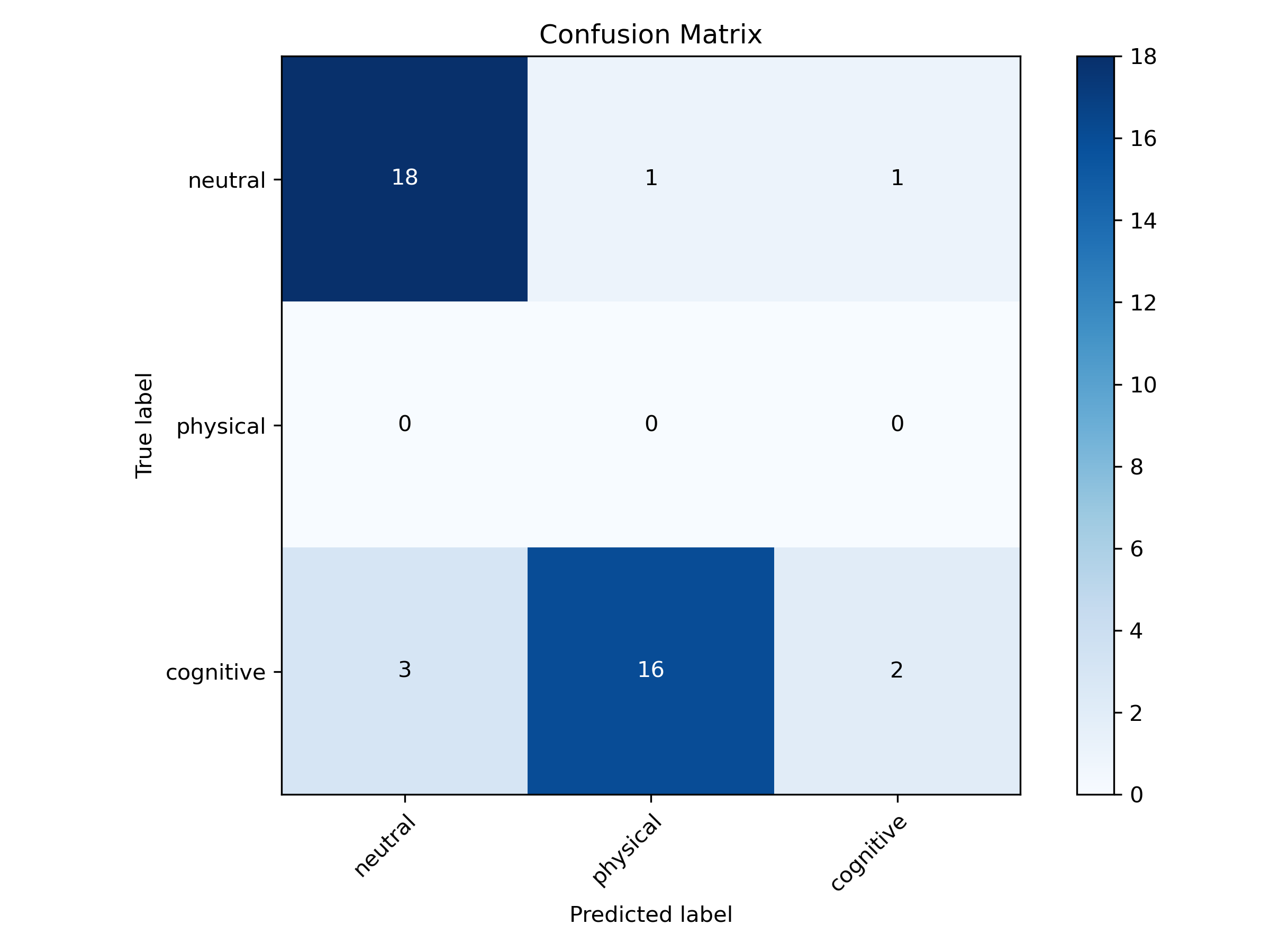}
  \end{minipage}\hfill
  \begin{minipage}{0.3\textwidth}
    \centering
    \includegraphics[width=\linewidth]{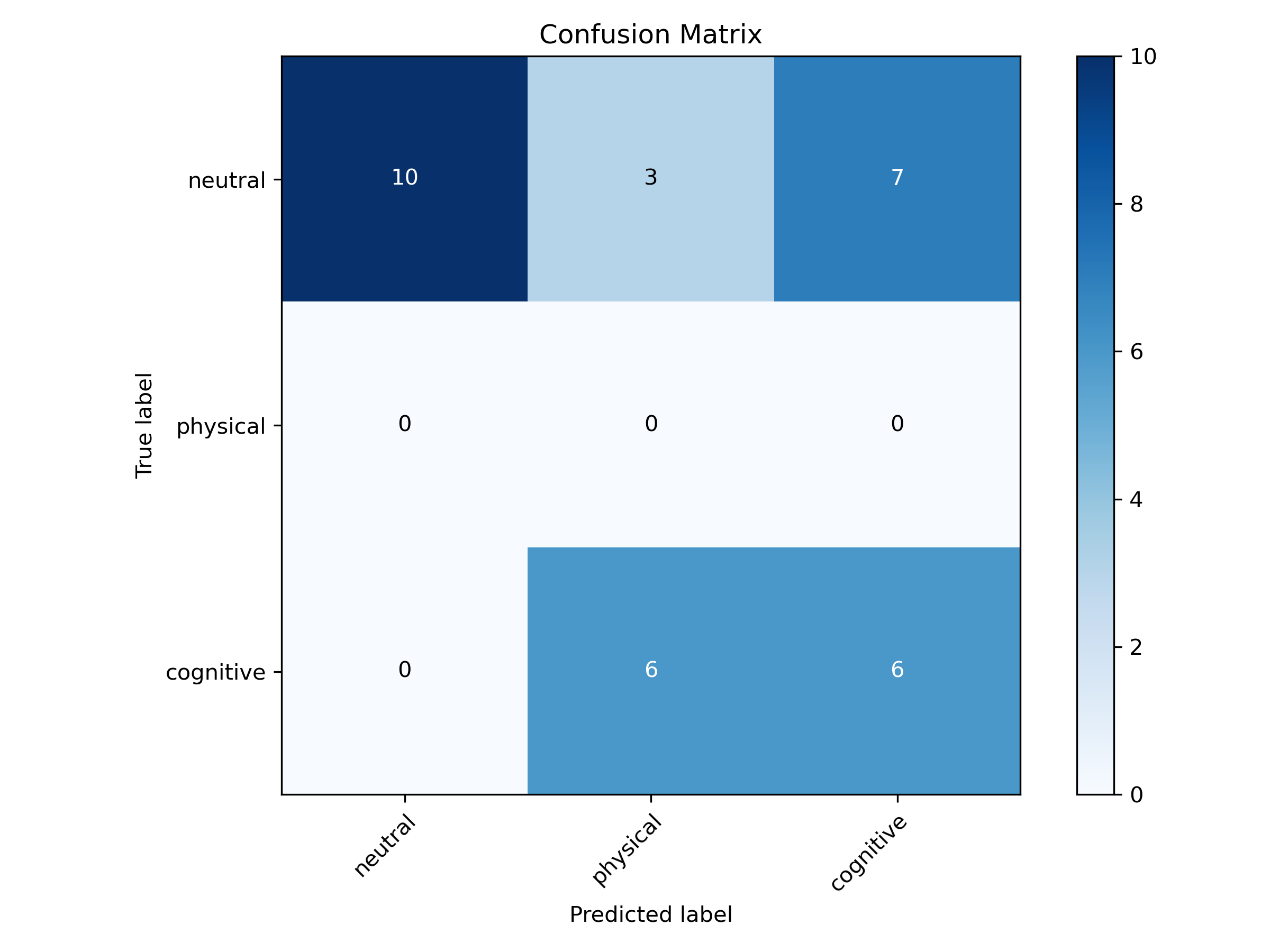}
  \end{minipage}\hfill
  \begin{minipage}{0.3\textwidth}
    \centering
    \includegraphics[width=\linewidth]{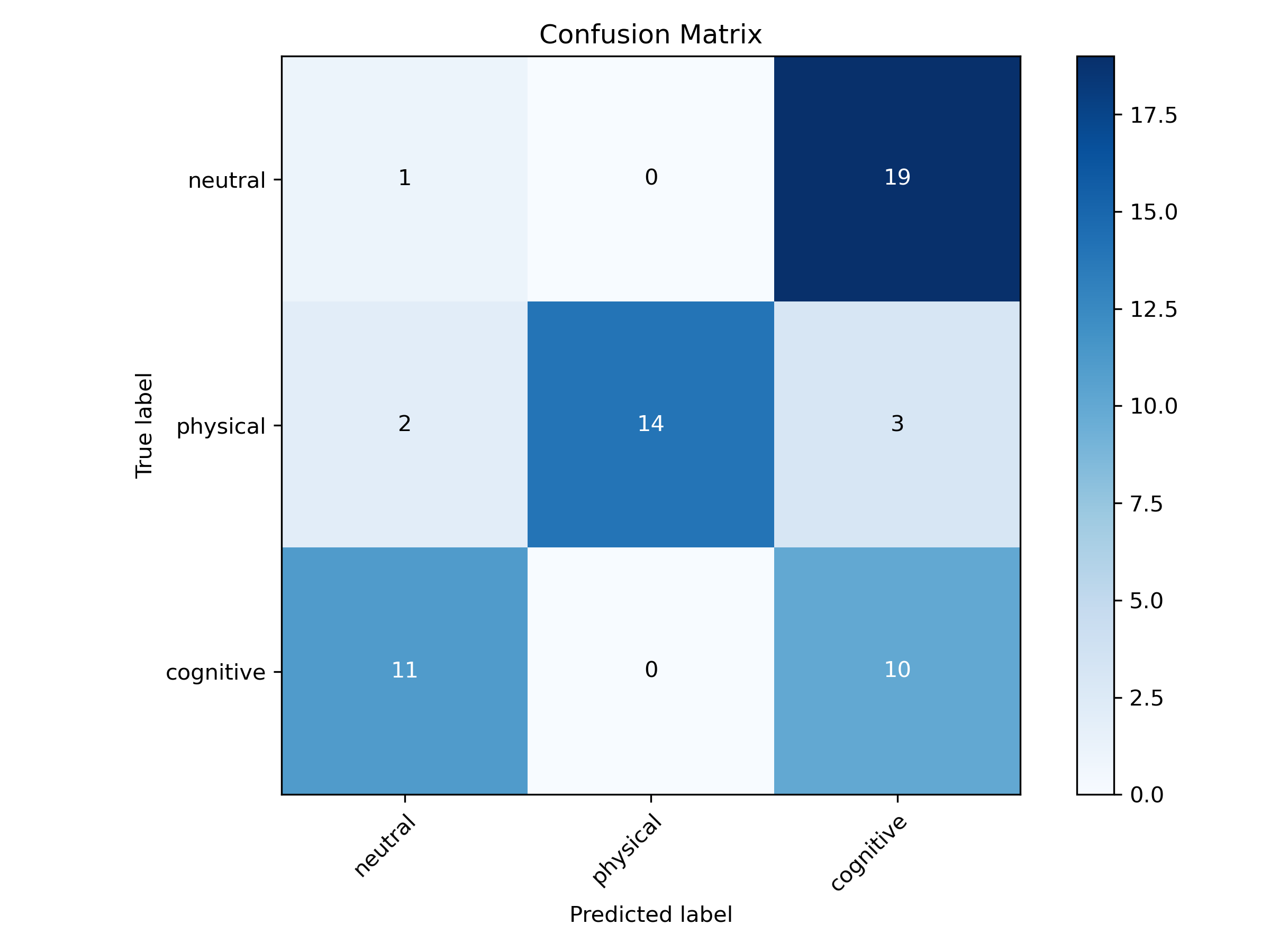}
  \end{minipage}
  \caption{Mis-classification patterns.}
  \label{fig:combined_confusion_matrices}
\end{figure}

Upon further evaluation of the model performance in the ternary classification task, we observe that for some test subjects, the models did not generalize well, resulting in poor classification performance in terms of accuracy. For 3 out of 21 test subjects, the model achieved less than 64\% in classification accuracy. Figure \ref{fig:combined_confusion_matrices} provide examples of test subjects, showing the classification patterns for three subjects with low accuracy. These results indicate that the globally trained model performed well for most test subjects. However, these results also suggest that global models, rather than subject-specific ones, do not always learn the subject specific intricacies.

The results presented in study make apparent that the motion features are very informative for the classifiers, as when removed the classification results are on average 3\% worse compared to when all features are available. The motion features, however, have been subject to the most pre-processing and aggregation, such that feature engineering of the motion features may have been very efficient.

As discussed before, the results reported here are the averaged accuracies and F1-scores from the LOSO-CV procedure applied during training. An inspection of the results of the different test splits showed that for most subjects, the model was able to generalize well beyond the training data. This insight also revealed that the models performed very poorly on some subjects. However, a detailed analysis of the errors or bad generalization capabilities for some subjects is out of the scope of this paper.

\section{Discussion}
\label{sec:discussion}

The preliminary analysis of features presented in this research distinguishes neutral baseline, cognitive stress, socio-evaluative stress, and physical exertion. The violin plots show that physical activity produces the most significant accelerometer variability and heart-rate elevations, cognitive load yields minimal motion with pronounced electrodermal responses, and public speaking occupies an intermediate profile across all measures. The principal component analysis reinforces this pattern: PC1 (40.7\% of variance) captures a "motion-cardiac" axis driven by accelerometer standard deviations and heart-rate (bpm) - on which physical exertion scores highest - while PC2 (10\%) defines a "posture-stress" axis contrasting vertical bias and electrodermal activity against lateral movement and thermal and cardiac signals, thus separating static low-stress (neutral baseline) from static high-stress (cognitive) conditions. These results imply that a limited but compact feature set of physiological and kinematic signals can reliably discriminate among diverse physiological and psychological states in real time.

The collected dataset was employed to distinguish between the presence and absence of stress and to differentiate between different types of stress states (physical, cognitive, socio-evaluative). The use cases showed that simple classifiers reliably detect stressful events, the models reported promising outcomes which can be further improved or used for automated stress recognition systems. The results presented in this paper are similar to the results presented in previous studies on this topic (e.g. \citep{schmidt2018introducing, hovsepian2015cstress}.

Beyond evaluating algorithmic performance on automated stress recognition, the dataset's structure and rigorous annotation enable the exploration of various downstream tasks. The dataset allows for the benchmarking of new classification models, the development of subject-specific stress prediction systems, and the evaluation of generalization across individuals. Additionally, it provides a foundation for investigating feature importance across modalities (e.g., heart-rate vs movement data) and validating real-world stress detection scenarios outside of laboratory settings. Use cases of the multi-modal VitaStress dataset include but are not limited to the development of automated stress detection models and affective user modeling. For instance, developing generalizable stress recognition systems requires data integration from multiple sources. As such, the \textit{VitaStress} dataset supports current research in affective computing and wearable health technology and promotes reproducibility and comparability -  critical factors for long-term impact in human-centered machine learning.

To the best of our knowledge, this dataset is the first to combine cognitive and socio-evaluative stress with physical stress, i.e., physical exercise, while providing explicit labels for each stressor. \cite{birjandtalab2016non} contributed a similar dataset; however, the  VitaStress dataset includes detailed information about the procedures of the experiments, such that labels can be inferred from the data. Other studies (e.g. \citep{mishra2020continuous, plarre2011continuous}) employ a cold-pressor task as physical stress; however, as \cite{mishra2020continuous} pointed out, the cold-pressor stimuli might not lead to significantly different physiological responses than the neutral baseline.

Integrating various data sources is a major challenge in automated stress recognition and user modeling. Stress recognition systems involving machine- and deep learning architectures would vastly benefit from increased volume of training samples. As shown in section \ref{sec:related_work}, datasets often differ drastically, decreasing their comparability. Standardizing entire experimental protocols, rather than just stimuli, would facilitate the structural unity among datasets.

\section{Conclusion}
\label{sec:conclusion}

We present VitaStress, a multi-modal dataset for developing automated stress recognition systems. To the best of our knowledge, this dataset is the first of its kind, including cognitive stress, socio-evaluative stress, and physical stress with explicit labeling and rigorous annotations. We show that a combination of vital and motion data can be reliably employed to detect stress with the help of machine learning methods. In particular, it is highlighted that multi-modal data enhance performance and that motion data is crucial to detect physical stress. The analysis results further demonstrate that subject-independent modeling yields good performances for the majority of the test subjects. However, the models evaluated here fail to generalize well for all test subjects, indicating the necessity of individualization for deploying automated stress recognition systems in real-life scenarios. This study classified between neutral baseline, physical and cognitive (cognitive and social) stress. The comparison to binary classification showed that both approaches are feasible while binary classification leads to better results since the model sometimes fails to differentiate cognitive stress (cognitive and social) from neutral and physical stress. Furthermore, by collecting empirical data, we validate the data collection and re-usability framework TRRRACED. The systematic review provides a comprehensive overview of currently publicly available datasets, which sample physiological data in laboratory settings.

As part of ongoing work, we are conducting a more detailed analysis of the data available in this dataset. As mentioned before, the results of this and prior work have shown that compact, selected features reliably distinguish between the different conditions. Yet, we anticipate that including raw data, sampled at a much higher frequency, will refine our analysis. Finally, we aim to expand the dataset to gather more samples and increase the demographic variability of our samples.

\bibliographystyle{unsrtnat}
\bibliography{references}  %%% Uncomment this line and comment out the ``thebibliography'' section below to use the external .bib file (using bibtex) .





% \end{thebibliography}

\end{document}